\documentclass[twocolumn]{aastex701}
\usepackage{amsmath}
\usepackage{makecell}
\usepackage{multirow}
\usepackage{colortbl}
\usepackage{xcolor}
\usepackage{booktabs}

\definecolor{darkblue}{RGB}{200,55,0}
\definecolor{darkred}{RGB}{0,200,55}

\begin{document}

\title{Seeing Through the RFI Mask: DT-FISTA Inpainting of Pulsar Dynamic Spectra via Fourier Sparsity}

\author[0009-0006-4956-8803]{Yuanshang Huang}
\affiliation{South-Western Institute for Astronomy Research (SWIFAR), Yunnan University, Kunming 650500, China}
\email{huangyuanshang42@gmail.com}
\author[0000-0003-2076-4510]{Xun Shi}
\affiliation{South-Western Institute for Astronomy Research (SWIFAR), Yunnan University, Kunming 650500, China}
\email{xun@ynu.edu.cn}

\begin{abstract}

Radio frequency interference (RFI) frequently contaminates radio observations, and excising affected samples leaves masked regions in sampled images. Replacing masked samples with zero or the mean causes spectral leakage in Fourier-domain power representations. This contamination is particularly severe for modern pulsar scintillation studies, which rely on identifying and characterizing faint parabolic scintillation arcs in secondary spectra, the 2D power spectra of dynamic spectra. Scintillation arcs probe discrete scattering screens in the interstellar medium (ISM) and constrain their physical properties. However, leakage can obscure faint arcs whose power is often $<10^{-3}$ of the dominant pulsar signal power. Wiener inpainting can mitigate this leakage, but a dense implementation requires $\mathcal{O}(N^2)$ memory and $\mathcal{O}(JN^2)$ time for $N$ pixels and $J$ iterations, which is computationally expensive for full-resolution spectra containing millions of pixels. We introduce the decreasing-threshold fast iterative shrinkage-thresholding algorithm (DT-FISTA). Using a sparse Fourier representation, DT-FISTA initially estimates masked intensities from the strongest Fourier components and progressively refines them with weaker components as the soft threshold decreases. Nesterov momentum extrapolation accelerates these updates. Each iteration uses 2D fast Fourier transforms (FFT) and pointwise operations, requiring $\mathcal{O}(N\log N)$ time and $\mathcal{O}(N)$ memory. Tests on $300\times300$ pixel patches from Five-hundred-meter Aperture Spherical radio Telescope (FAST) observations yield lower reconstruction errors than the dense Wiener implementation, with runtime and memory requirements reduced by factors $>10^3$. Applied to full-resolution dynamic spectra, DT-FISTA suppresses mask-induced artifacts and reveals previously obscured faint arcs, enabling efficient ISM studies.

\end{abstract}


\section{Introduction} \label{sec:intro}

Radio waves from pulsars are scattered by electron density irregularities in the ionized interstellar medium (ISM), which are often modeled as localized thin scattering screens (e.g., \citealt{rickett90}). As the pulsar, Earth, and scattering screen move, interference among different propagation paths evolves with time and frequency, modulating the received intensity and producing interstellar scintillation (e.g., \citealt{rickett90}). The dynamic spectrum records this intensity as a function of time and observing frequency, and the squared modulus of its two-dimensional Fourier transform is known as the secondary spectrum. In the secondary spectrum, power is often concentrated along parabolic scintillation arcs \citep{stinebring01} and smaller inverted parabolic arclets \citep{hill2005}, both of which arise from pairwise interference among distinct scattered rays. The relative transverse motion of the pulsar, scattering screen, and observer sets the differential Doppler coordinate, while the differential geometric delay depends on the difference between squared scattering angles, producing the observed parabolic loci \citep{walker04,cordes06,main23}. The arc curvature and substructure constrain the screen location, transverse velocity, and scattering anisotropy, while observations at multiple epochs or additional geometric information can help break degeneracies among these parameters \citep{reardon20}.

During the observation of a dynamic spectrum, various factors can prevent the signal from being robustly measured at specific coordinates in the time-frequency domain, leading to missing data. Most notably, RFI mitigation pipelines flag and excise contaminated samples, leaving data gaps in the dynamic spectrum \citep{fridman01, offringa10, offringa12}. A simple treatment is to replace masked samples with zero or with the mean of the retained data. In this case, the observed dynamic spectrum becomes the underlying intensity multiplied by a binary sampling mask. By the convolution theorem, this results in the convolution of the signal's Fourier transform with that of the mask in the conjugate domain. The resulting sidelobes redistribute power from bright scintillation features in the secondary spectrum, producing spectral leakage that can obscure faint, extended arcs. Inpainting the masked signals mitigates this leakage.

Wiener filtering \citep{Wiener_1949} is an established approach for inpainting masked samples in pulsar dynamic spectra. In the formulation of \citet{lin21}, an estimated signal covariance and a regularizing noise term determine a linear estimator for the masked samples, an approach they applied to a masked pulsar eclipse. Direct inversion of the dense $N\times N$ Wiener covariance matrix requires $\mathcal{O}(N^3)$ time and $\mathcal{O}(N^2)$ memory; solving the system with $J$ conjugate-gradient iterations reduces the time cost to $\mathcal{O}(JN^2)$. This prohibitive computational cost has previously led authors like \citet{main23} and \citet{reardon25} to divide observations into small blocks of $N \sim 100^2$ and apply Wiener inpainting separately to each. Furthermore, the accuracy of Wiener inpainting relies on how accurately the estimated signal and noise covariances describe the data.

Fourier sparsity provides an alternative approach for inpainting masked samples without constructing a dense covariance matrix. Compressed sensing theory guarantees accurate signal recovery when a signal is sparse or compressible in a known representation, provided the sampling pattern supplies sufficient information \citep{candes06, donoho06}. For pulsar dynamic spectra, the concentration of power along narrow arcs motivates an approximate sparsity prior in the Fourier representation. Mathematically, this can be formulated via constrained $\ell_1$ norm minimization, commonly known as basis pursuit \citep{chen98}, while iterative thresholding provides algorithms for solving such sparsity-regularized inverse problems \citep{daubechies04}.

Sparse and sparsity-based reconstruction methods have a long history in astronomy. A classical example is the Högbom CLEAN algorithm, a greedy iterative deconvolution method that can be interpreted as imposing an implicit point source sparsity prior \citep{hogbom74,wiaux09}. Later work formulated interferometric imaging as an explicit compressed-sensing inverse problem with sparsity regularization \citep{wiaux09}. Related sparse reconstruction methods have also been applied to asteroseismic time series inpainting \citep{pires15}. In pulsar studies, \citet{oslowski23} introduced H-FISTA, a hierarchical phase-retrieval method based on FISTA for pulsar dynamic spectra. H-FISTA provides a related precedent for sparse reconstruction of the wavefield underlying pulsar dynamic spectra.

We propose DT-FISTA to inpaint masked regions by progressively refining a dynamic spectrum estimate via FFTs. Initially, an inverse FFT of the strongest Fourier components after soft shrinkage provides a coarse estimate of the masked samples. As iterations proceed, the threshold decreases, weaker components are admitted, progressively adding fine detail to the masked regions. This strategy is motivated by the mathematical concept of continuation \citep{hale08}. Furthermore, Nesterov momentum extrapolation uses the difference between successive iterations to accelerate convergence \citep{nesterov1983, beck2009}. Computationally, each iteration requires only one forward and one inverse two-dimensional FFT and pointwise operations, meaning $K$ iterations scale with $\mathcal{O}(KN\log N)$ time and $\mathcal{O}(N)$ memory. We demonstrate this method on full-resolution FAST dynamic spectra, quantifying both inpainting accuracy and computational efficiency.

\begin{figure*}[tp]
\centering
\includegraphics[width=0.95\textwidth]{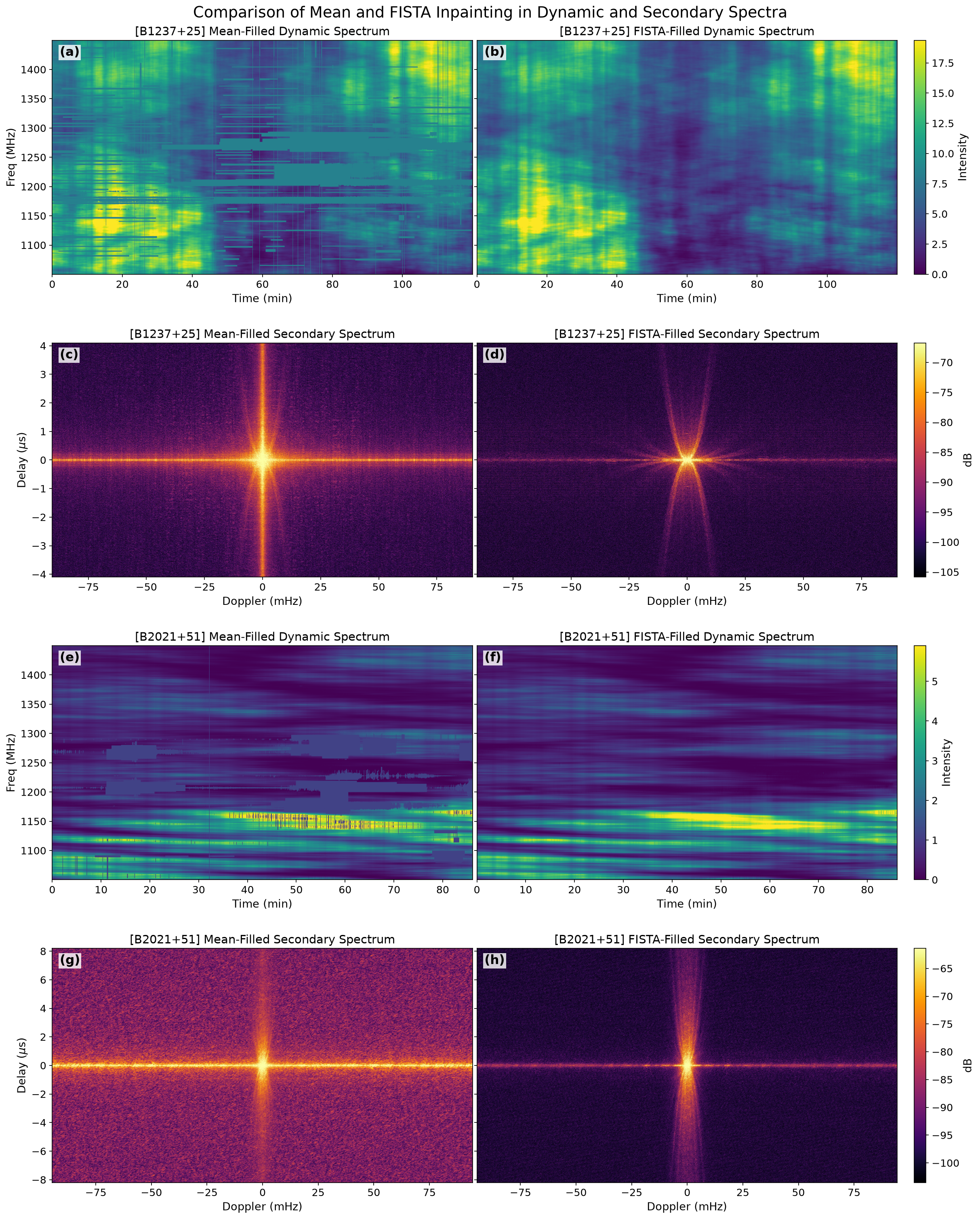}
\caption{Inpainting at full-resolution suppresses mask-induced leakage in the secondary spectrum and makes faint scintillation arcs more visible. The dynamic and secondary spectra of PSR B1237+25 and PSR B2021+51 are shown after mean value inpainting (left panels) and DT-FISTA (right panels). Secondary spectrum power is displayed in dB, with the power at zero delay and zero Doppler (DC) normalized to 0 dB.}
\label{fig:full_reconstruction}
\end{figure*}

Figure~\ref{fig:full_reconstruction} illustrates this problem in full-resolution observations. Mean-value filling produces mask-induced leakage in the secondary spectra, whereas DT-FISTA suppresses the leakage and makes faint scintillation arcs more visible.

The remainder of this paper is organized as follows. Section~\ref{sec:method} describes the method. Section~\ref{sec:results} presents the performance of DT-FISTA in the validation tests. Section~\ref{sec:conclusion} summarizes the main results and conclusions.

\section{Method: Inpainting Masked Samples from Strong to Weak Fourier Structure} \label{sec:method}

\subsection{Algorithm Overview} \label{subsec:overview}

The central concept is to inpaint masked values using dominant Fourier components first, and then progressively refine these estimates with weaker components as the shrinkage threshold decreases.

The inputs are a dynamic spectrum and a binary mask. The threshold starts at a large Fourier amplitude. Each iteration applies the following five operations:

\begin{enumerate}

\item \textbf{Fourier transformation.} The current dynamic spectrum estimate is transformed into the two-dimensional Fourier domain.

\item \textbf{Soft shrinkage.} Soft thresholding is applied to all non-DC coefficients. Magnitudes below the threshold are set to zero, while larger magnitudes are reduced by the threshold with phases preserved. The DC coefficient remains unchanged.

\item \textbf{Inverse Fourier transformation.} The shrunken coefficients are transformed back to the time-frequency domain, yielding a new estimate over the full grid.

\item \textbf{Data consistency.} Retained samples are restored to their exact input values from the preprocessed dynamic spectrum, while masked samples keep the new inverse transform estimates.

\item \textbf{Momentum extrapolation.} The difference between the two latest estimates is multiplied by a weight and added to the newest estimate to form the input for the next iteration. The weight follows the standard FISTA recurrence \citep{nesterov1983, beck2009}, with the first update having zero momentum.

\end{enumerate}

After each cycle, the threshold is multiplied by a decay factor less than unity.

\begin{figure*}[tp]
\centering
\includegraphics[width=\textwidth]{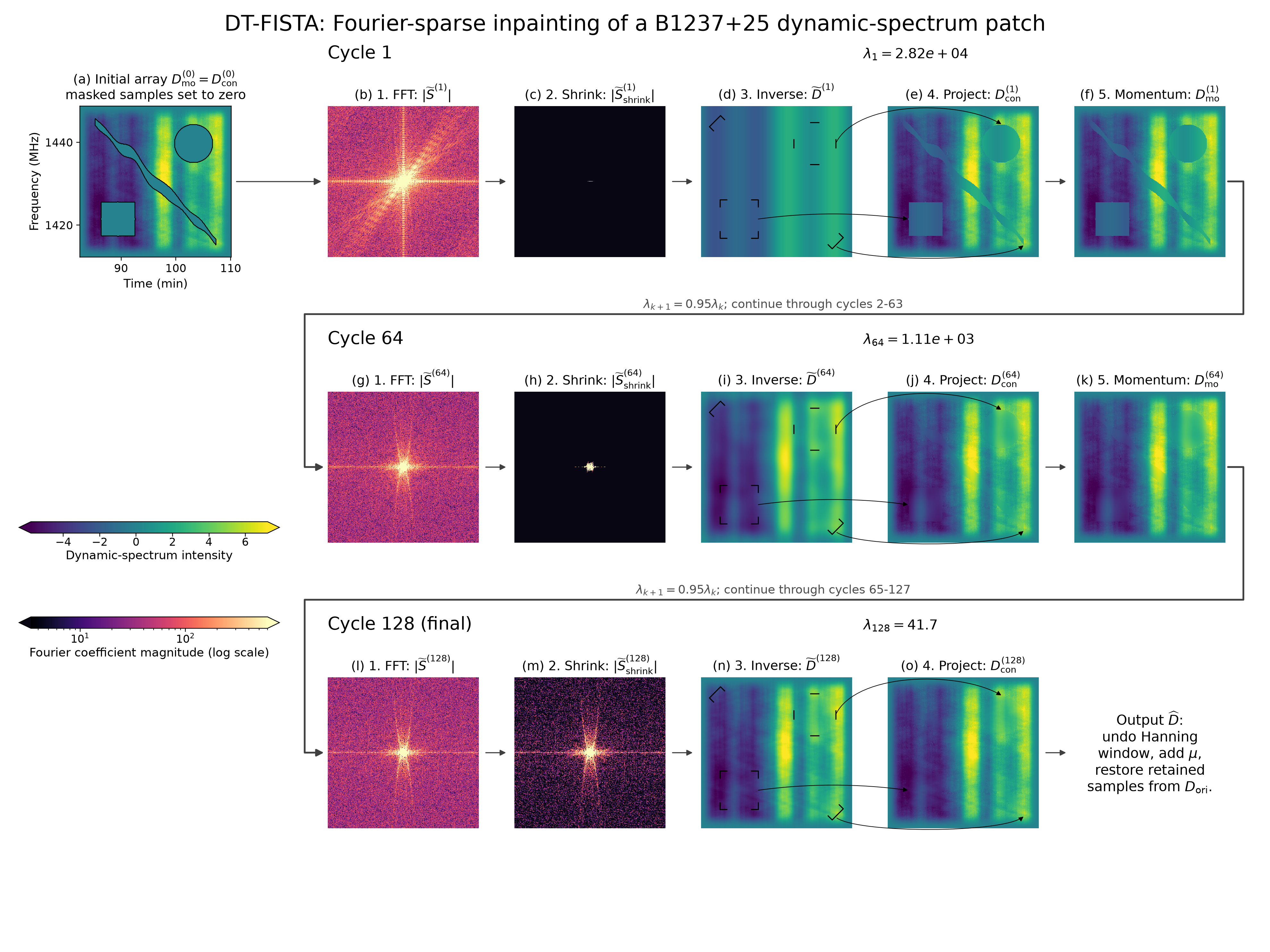}
\caption{Flowchart of DT-FISTA inpainting for a $300\times300$ patch of the pulsar dynamic spectrum of B1237+25. Panel (a) shows the initial array $D_{\mathrm{mo}}^{(0)}=D_{\mathrm{con}}^{(0)}$, whose masked samples are set to zero. The regions enclosed by solid black outlines in panel (a) are the masked regions. Each cycle $k$ consists of five operations: Fourier transformation of the momentum input $D_{\mathrm{mo}}^{(k-1)}$ to obtain $\widetilde{S}^{(k)}$, complex soft shrinkage to obtain $\widetilde{S}_{\mathrm{shrink}}^{(k)}$, inverse transformation to obtain $\widetilde{D}^{(k)}$, data consistency projection to obtain $D_{\mathrm{con}}^{(k)}$, and FISTA momentum extrapolation to obtain $D_{\mathrm{mo}}^{(k)}$. The three rows show cycles 1, 64, and 128, with $\lambda_{k+1}=0.95\lambda_k$ between cycles. Arrows between the inverse transform and projection panels indicate the estimates inserted into the masked regions. The final output $\widehat{D}$ is obtained by undoing the Hanning window, restoring the mean, and copying the retained samples from $D_{\mathrm{ori}}$. The Fourier panels display coefficient magnitudes on a logarithmic scale, with low Fourier frequencies at the center and high Fourier frequencies toward the edges.}
\label{fig:confista_flowchart}
\end{figure*}

\subsection{Sparse Inpainting Constraint} \label{subsec:formulation}

Let $D_{\mathrm{ori}} \in \mathbb{R}^{N_\nu \times N_t}$ denote the original dynamic spectrum, spanning $N_\nu$ frequency channels and $N_t$ time samples, and let $M \in \{0, 1\}^{N_\nu \times N_t}$ be the binary sampling mask, with $M=1$ at retained samples and $M=0$ at masked samples. The mean $\mu$ is computed from the retained samples. A Hanning window $W$ tapers the outer 10\% at each end of both axes and equals unity in the interior. The mean subtracted, windowed dynamic spectrum is given by $D_{\mathrm{sub}\mu,\mathrm{han}} = W \odot (D_{\mathrm{ori}} - \mu)$ at retained samples. Its values in masked regions are set to zero to define the initial data constrained array $D_{\mathrm{con}}^{(0)}$, which acts as a mean-value inpainting prior before mean subtraction and windowing. Let $D \in \mathbb{R}^{N_\nu \times N_t}$ be an inpainted estimate, and let $\odot$ denote element wise multiplication. The data consistency constraint set is defined as:

\begin{equation} \label{eq:cdata}
\mathcal{C}_{\mathrm{data}}=\{D:M\odot D=M\odot D_{\mathrm{con}}^{(0)}\}.
\end{equation}
Enforcing this data consistency is equivalent to an orthogonal projection onto a convex set, as used in the projection onto convex sets (POCS) framework \citep{youla82, sezan82}. The Euclidean projection of any dynamic spectrum $D_*$ onto this set is:
\begin{equation} \label{eq:pdata}
\mathcal{P}_{\mathrm{data}}(D_*)=M\odot D_{\mathrm{con}}^{(0)}+(1-M)\odot D_*.
\end{equation}
Equivalently, this projection operates pixel by pixel as:
\begin{equation} \label{eq:pdata_elementwise}
\mathcal{P}_{\mathrm{data}}(D_*)_{i,j}=\begin{cases}
D_{\mathrm{con},i,j}^{(0)},&M_{i,j}=1,\\
D_{*,i,j},&M_{i,j}=0.
\end{cases}
\end{equation}

A reference problem based on basis pursuit mathematically formalizes the preference for Fourier sparsity:

\begin{equation} \label{eq:bp_reference}
\widehat{D}_{\mathrm{BP}}\in\arg\min_{D\in\mathcal{C}_{\mathrm{data}}}\mathcal{R}(D),
\quad
\mathcal{R}(D)=\sum_{\omega\ne(0,0)}|\mathcal{F}(D)_\omega|.
\end{equation}

Here, $\mathcal{F}$ represents the two-dimensional FFT, $\omega=(\omega_\nu,\omega_t)$ indexes the delay and Doppler coordinates, and ``BP'' denotes a basis pursuit type problem. We distinguish the complex Fourier coefficients $\widetilde{S}(D)=\mathcal{F}(D)$ from the secondary spectrum $S(D)=|\widetilde{S}(D)|^2$. The penalty $\mathcal{R}$ and the data consistency set $\mathcal{C}_{\mathrm{data}}$ are both convex, as proven in Appendix~\ref{app:convexity}.

While Equation~(\ref{eq:bp_reference}) acts as the theoretical objective, it is optimized implicitly within the DT-FISTA framework. The non-DC $\ell_1$ norm $\mathcal{R}$ serves as the sparsity prior favoring a compact Fourier representation, which is algorithmically enforced via its proximal operator (the soft shrinkage in Step 2). Concurrently, the constraint $\mathcal{C}_{\mathrm{data}}$ is strictly satisfied through exact projection (Step 4). The exact mathematical correspondence between these operations and standard proximal gradient methods is detailed in Section~\ref{subsec:standard_fista}.

Fourier sparsity naturally guides the inference of masked pixels. Because the retained samples only constrain entries outside the masks, any values placed within the masked regions, including zeros, perfectly reproduce the observed data. Consequently, choosing among these infinitely many consistent arrays requires a prior. The sparse inpainting prior, modeled by the $\ell_1$ penalty, favors reconstructions that concentrate the Fourier representation, exploiting the inherent compressibility of the noisy reference signal.

\subsection{The DT-FISTA Algorithm} \label{subsec:confista}

The algorithm iterates through the five operations outlined in Section~\ref{subsec:overview}, with the first iteration illustrated in Figure~\ref{fig:confista_flowchart} (b)-(f). Starting from the mean-value initialization $D_{\mathrm{con}}^{(0)}$, we initialize:

\begin{equation} \label{eq:init}
D_{\mathrm{mo}}^{(0)}=D_{\mathrm{con}}^{(0)},\qquad t_0=1.
\end{equation}

Here, $D_{\mathrm{con}}^{(k)}$ is the result of the data-consistency projection and $D_{\mathrm{mo}}^{(k)}$ is the momentum extrapolated array produced at the end of cycle $k$ for use in cycle $k+1$. Superscript $(0)$ denotes the initial state; all arrays updated within cycle $k$ carry superscript $(k)$. For $k=1,2,\ldots$, each cycle proceeds as follows:

\noindent\textbf{1. Fourier transformation.} Transform the momentum input into the Fourier domain:
\begin{equation} \label{eq:fft}
\widetilde{S}^{(k)}=\mathcal{F}\!\left(D_{\mathrm{mo}}^{(k-1)}\right).
\end{equation}
The corresponding secondary spectrum is $S^{(k)}=|\widetilde{S}^{(k)}|^2$.

\noindent\textbf{2. Soft shrinkage.} Apply complex soft thresholding to $\widetilde{S}^{(k)}$:

\begin{equation} \label{eq:soft}
\widetilde{S}_{\mathrm{shrink},\omega}^{(k)}=
\max\!\left(|\widetilde{S}^{(k)}_\omega|-\lambda_k,0\right)
\frac{\widetilde{S}^{(k)}_\omega}{|\widetilde{S}^{(k)}_\omega|},\quad \omega\ne(0,0),
\end{equation}

yielding zero when $\widetilde{S}^{(k)}_\omega=0$, and preserving the DC component identically: $\widetilde{S}_{\mathrm{shrink},(0,0)}^{(k)}=\widetilde{S}^{(k)}_{(0,0)}$. Coefficients at or below the threshold vanish, while surviving coefficients have their magnitude reduced by $\lambda_k$ without altering their phase. Preserving the DC component ensures that the mean of the inpainted array can adapt naturally as masked samples are inferred. The secondary spectrum after shrinkage is $S_{\mathrm{shrink}}^{(k)}=|\widetilde{S}_{\mathrm{shrink}}^{(k)}|^2$.

\noindent\textbf{3. Inverse Fourier transformation.} Transform the shrunken coefficients back and retain the real part:
\begin{equation} \label{eq:ifft}
\widetilde{D}^{(k)}=\operatorname{Re}\!\left\{\mathcal{F}^{-1}\!\left(\widetilde{S}_{\mathrm{shrink}}^{(k)}\right)\right\}.
\end{equation}
Because the soft shrinkage is applied symmetrically to the magnitudes of conjugate pairs, it analytically preserves the Hermitian symmetry of the Fourier array for real valued inputs; thus, any imaginary residue following the inverse transform is purely numerical roundoff.

\noindent\textbf{4. Data consistency projection.} Restore the retained samples to their exact input values, so that only the masked pixels are updated:

\begin{equation} \label{eq:replacement}
D_{\mathrm{con},i,j}^{(k)}=\begin{cases}
D_{\mathrm{con},i,j}^{(0)},&M_{i,j}=1,\\
\widetilde{D}^{(k)}_{i,j},&M_{i,j}=0.
\end{cases}
\end{equation}

\noindent\textbf{5. Momentum extrapolation.} Combine the projected dynamic spectrum from the current cycle with that from the preceding cycle to formulate the momentum input:

\begin{equation} \label{eq:confista_t}
t_k=\frac{1+\sqrt{1+4t_{k-1}^2}}{2},\qquad
\beta_k=\frac{t_{k-1}-1}{t_k},
\end{equation}
\begin{equation} \label{eq:confista_v}
D_{\mathrm{mo}}^{(k)}=D_{\mathrm{con}}^{(k)}+\beta_k\left(D_{\mathrm{con}}^{(k)}-D_{\mathrm{con}}^{(k-1)}\right).
\end{equation}

Because the difference between two projected spectra is identically zero at every retained pixel, momentum alters only the masked pixels. The first update yields $\beta_1=0$, meaning $D_{\mathrm{mo}}^{(1)}=D_{\mathrm{con}}^{(1)}$, as depicted in Figure~\ref{fig:confista_flowchart} (e, f). At iteration 64, $\beta_{64}\simeq0.956$ extrapolates the estimate in Figure~\ref{fig:confista_flowchart} (j) to the subsequent input in Figure~\ref{fig:confista_flowchart} (k).

This strategy addresses the problem of mask-induced spectral leakage through a coarse-to-fine progression. Initially, the sharp discontinuities of the data gaps scatter artificial sidelobes across the Fourier domain. A high initial threshold restricts early updates to dominant signal peaks, suppressing most low-amplitude leakage. As the algorithm populates the masked regions with these reliable strong components, the sharp edges in the dynamic spectrum are smoothed out. With the artifact background diminished, the algorithm can sequentially lower the threshold to capture weaker, finer Fourier components while minimizing the inclusion of remaining leakage. This progressive refinement is visible in the soft shrinkage sequence (Figure~\ref{fig:confista_flowchart} c, h, m) and the increasingly detailed estimates (Figure~\ref{fig:confista_flowchart} d, i, n).

To implement this threshold decay, we use
\begin{equation} \label{eq:decay} \lambda_{k+1}=\kappa\lambda_k,\qquad 0<\kappa<1,
\end{equation}

with $\kappa=0.95$ adopted as the default for the reported tests. Empirical validation indicates that inpainting accuracy is highly robust to $\kappa \in [0.9, 0.99]$, though values closer to unity require proportionally more iterations. The initial threshold, $\lambda_1$, is conservatively set to the tenth largest value of $|\widetilde{S}^{(1)}|$. Starting with such a high threshold rigorously restricts early updates to dominant signal peaks.

At the start of each iteration, we evaluate the median amplitude:

\begin{equation} \label{eq:median_scale}
m_k=\operatorname{median}(|\widetilde{S}^{(k)}|).
\end{equation}

Iterations proceed as long as $\lambda_k\ge m_k$ and terminate before shrinkage at the first occurrence of $\lambda_k<m_k$. If this happens at cycle $k$, then $K=k-1$ full cycles have completed, and $D_{\mathrm{con}}^{(K)}$ serves as the final internal state. Here, $m_k$ provides a robust, data driven estimate of the baseline Fourier noise level (assessed empirically in Section~\ref{subsec:patch_results}). Finally, dividing by the window and adding the mean:

\begin{equation} \label{eq:output}
\widehat{D}_{i,j}=\begin{cases}
D_{\mathrm{ori},i,j},&M_{i,j}=1,\\
D_{\mathrm{con},i,j}^{(K)}/W_{i,j}+\mu,&M_{i,j}=0.
\end{cases}
\end{equation}

We extract the final output from \(D_{\mathrm{con}}^{(K)}\) rather than \(D_{\mathrm{mo}}^{(K)}\) because the former directly reflects the noise suppressed shrinkage of the final iteration, whereas the latter contains un-shrunken extrapolation terms intended for a subsequent step. Note that dividing by small window values at the extreme edges can inadvertently amplify localized errors; a targeted edge repair procedure (Appendix~\ref{app:edge_repair}) resolves this boundary effect.

\subsection{Relation to Standard FISTA} \label{subsec:standard_fista}

DT-FISTA differs from the standard FISTA formulation in two ways: the ordering of proximal and gradient operations, and the dynamic threshold schedule. Consider the data fidelity objective function:

\begin{equation}
f(D) = \tfrac{1}{2} \|M \odot (D - D_{\mathrm{con}}^{(0)})\|_2^2,
\end{equation}

The analytical gradient of this objective is $\nabla f(D) = M \odot (D - D_{\mathrm{con}}^{(0)})$. Taking a gradient descent step with a unit step size replaces the retained entries with their exact known values from $D_{\mathrm{con}}^{(0)}$, leaving the masked entries untouched. Mathematically, this is identical to our data consistency projection: $D_* - \nabla f(D_*) = \mathcal{P}_{\mathrm{data}}(D_*)$.

In standard FISTA, a gradient step addressing data fidelity is explicitly followed by a proximal mapping step (soft shrinkage) addressing the $\ell_1$ penalty. In contrast, DT-FISTA reverses this order: it applies the proximal shrinkage before the gradient projection, ensuring that every completed cycle satisfies the data constraint $\mathcal{C}_{\mathrm{data}}$. Applying shrinkage first attenuates Fourier coefficients before the data-consistency projection and can help suppress low-amplitude noise and mask-induced leakage. The transition from $\widetilde{S}^{(64)}$ to $\widetilde{S}_{\mathrm{shrink}}^{(64)}$ (Figure \ref{fig:confista_flowchart} g, h) exemplifies this mechanism, where noise is truncated and signal amplitudes are conservatively scaled.

The second major difference is the threshold schedule. While standard FISTA assumes a fixed objective function with a static threshold, DT-FISTA actively decreases the threshold $\lambda_k$ across iterations. The name of the algorithm, Decreasing-Threshold FISTA, reflects this progressive relaxation of the $\ell_1$ penalty.

\section{Results} \label{sec:results}

\subsection{Full-resolution Inpainting Reveals Faint Arc Structure} \label{subsec:full_results}

DT-FISTA makes full-resolution inpainting practical for dynamic spectra containing millions of pixels. We applied mean-value inpainting and DT-FISTA to two FAST observations with several arcs: PSR~B1237+25 (MJD~58865.9) and PSR~B2021+51 (MJD~59839.5). These data were obtained from the FAST data archive. The dynamic spectra were constructed as described by \citet{huang2026}. We removed the lowest and highest 50~MHz of each observing band because they were recorded with different receivers, and manually flagged pixels affected by RFI. \citet{huang2026} identified 10 scintillation arcs toward PSR~B1237+25, including nine in MJD 58865.9 observation, and at least six toward PSR~B2021+51, all six visible in MJD 59839.5 observation. The arrays analyzed here contain $3276\times1304$ and $6552\times978$ pixels, respectively. We used $\kappa=0.95$ and the stopping rule based on the median Fourier magnitude in Section~\ref{subsec:confista}. The inpainted spectra in Figure~\ref{fig:full_reconstruction} include the second pass for edge repair described in Appendix~\ref{app:edge_repair}.

Suppressing mask-induced spectral leakage in the secondary spectrum exposes more of the arc structure. Figure~\ref{fig:full_reconstruction} compares the dynamic and secondary spectra. Mean-value inpainting leaves horizontal, vertical, and broad Fourier leakage. DT-FISTA suppresses these features, making faint arcs easier to distinguish from the background. The true values inside the observational RFI mask are unknown, so these full-resolution comparisons demonstrate improved visibility and reduced visible artifacts. Inpainting accuracy is measured with controlled masks below.

\subsection{Inpainting Test: Lower Error at Much Lower Cost} \label{subsec:patch_results}

We used patches of $300\times300$ pixels to compare mean-value inpainting, dense Wiener inpainting, and DT-FISTA inpainting so that the dense Wiener calculation remained tractable. A fully unmasked patch was available for PSR~B2021+51. No such patch was available for PSR~B1237+25, so we selected a patch with 677 originally masked pixels (0.75\%) and linearly interpolated those pixels before constructing the test. We then placed one square, one circular, and one irregular mask, each containing 4500 pixels (5\% of the patch), for a total of 13,500 masked pixels (15\% of the patch). The same mask geometry was used for both pulsars.

\subsubsection{Dense Wiener inpainting Used for Comparison} \label{subsec:wiener_details}

The Wiener inpainting follows the approach of Appendix~A of \citet{lin21}, with the noise term estimated from high frequency Fourier power as suggested therein. We inpaint the masked samples with the mean of the retained samples, subtract that mean, and apply the same Hanning window at the edges. The resulting values at masked pixels are zero. The stationary correlation kernel is obtained by taking the inverse Fourier transform of the secondary spectrum power and normalizing its zero lag to unity. The mean power in four corner regions, divided by the same normalization, supplies a white noise term. A dense covariance matrix is constructed from circular shifts of the correlation kernel, masked in both dimensions, and regularized with the noise term. The linear system is solved with unpreconditioned conjugate gradients (CG), using the default relative residual tolerance of $10^{-5}$ provided by SciPy \citep{2020SciPy-NMeth} in the tested environment. This tolerance measures the residual of the linear solve. A second covariance multiplication gives the Wiener estimate. We divide by the window, add the mean, and copy the estimates into the missing pixels.

The implementation constructs two 64-bit floating-point $N\times N$ matrices in sequence: one for the covariance solve and one for the final covariance multiplication. The first is released before the second is allocated. For a $300\times300$ patch, one dense matrix alone occupies 60.4 GiB, explaining the 60.4 GiB in Table \ref{tab:patch_metrics}. Matrix construction and the final multiplication cost $O(N^2)$, while $J$ conjugate gradient iterations cost $O(JN^2)$.

\begin{figure*}[tp]
\centering
\includegraphics[width=0.95\textwidth,height=0.89\textheight,keepaspectratio]{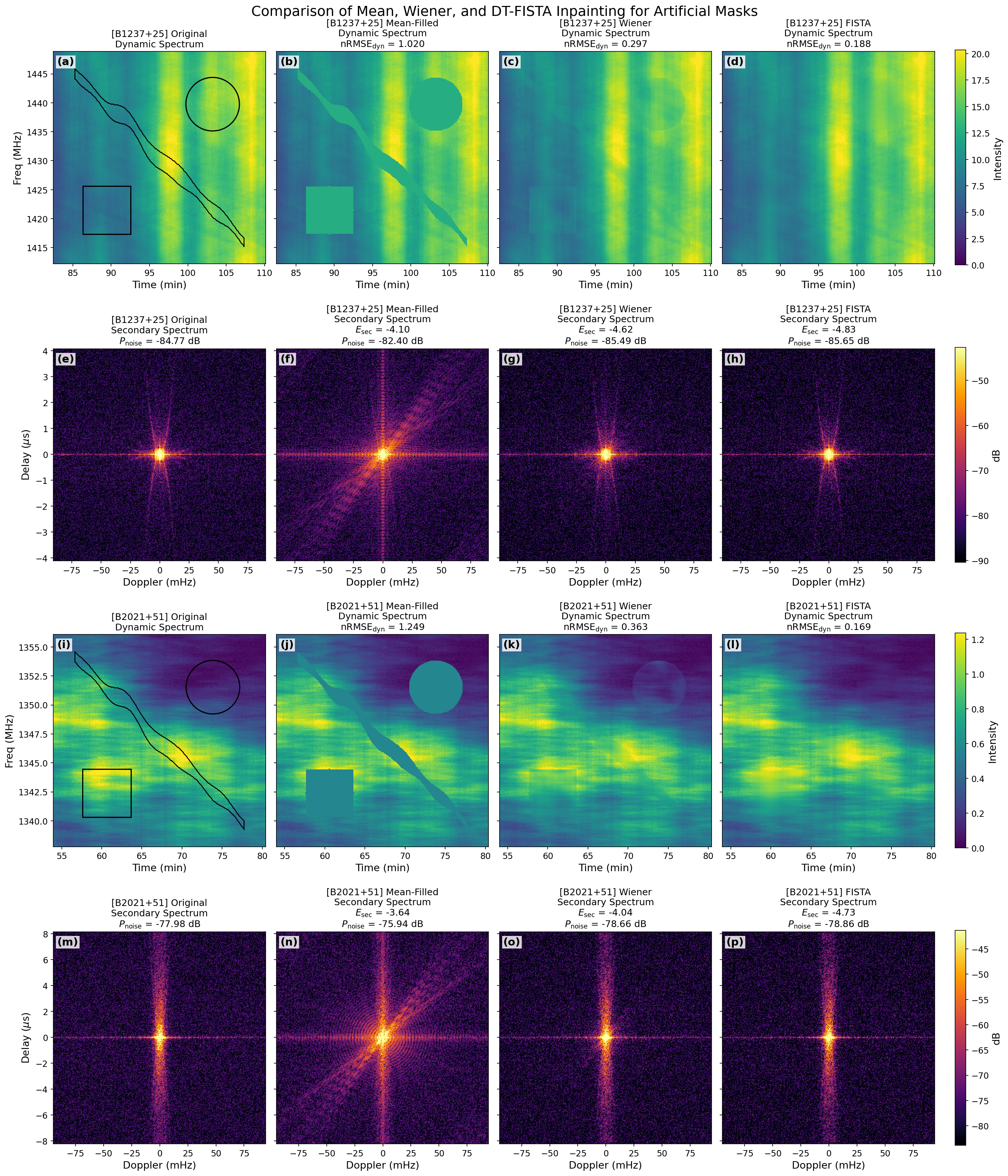}
\caption{DT-FISTA improves inpainting within controlled masks and the agreement of the resulting secondary spectrum with the reference. Tests use patches of $300\times300$ pixels. Solid black outlines in the original dynamic spectra delineate the mask regions. The square, circular, and irregular masks each have a nominal area equal to 5\% of the patch. The metrics $\mathrm{nRMSE}_{\mathrm{dyn}}$, $E_{\mathrm{sec}}$, and $P_{\mathrm{noise}}$ are defined in Section \ref{subsec:comparison}. Secondary spectrum power is in dB relative to DC.}
\label{fig:patch_comparison}
\end{figure*}

\subsubsection{Comparison} \label{subsec:comparison}

To quantitatively evaluate the inpainting accuracy in the controlled tests, we define three specific metrics. Let $D_{\mathrm{ori}}$ and $\widehat{D}$ denote the original and inpainted dynamic spectra of size $N_\nu \times N_t$, respectively. Let $M$ be the binary mask ($M_{i,j}=0$ for masked pixels), and let $\Omega_0 = \{(i,j) \mid M_{i,j}=0\}$ denote the set of missing pixels, with total count of $|\Omega_0|$. Their corresponding secondary spectra are $S_{\mathrm{ori}} = |\mathcal{F}(D_{\mathrm{ori}})|^2$ and $\widehat{S} = |\mathcal{F}(\widehat{D})|^2$, defined over the Fourier grid $\omega$.

\textbf{1. Dynamic spectrum normalized RMSE ($\text{nRMSE}_{\mathrm{dyn}}$):} 
We calculate the root-mean-square-error (RMSE) exclusively inside the artificially imposed mask, normalized by the standard deviation of the complete original patch, $\sigma(D_{\mathrm{ori}})$:
\begin{equation} \label{eq:metric_dyn}
\text{nRMSE}_{\mathrm{dyn}} = \frac{1}{\sigma(D_{\mathrm{ori}})} \sqrt{\frac{1}{|\Omega_0|} \sum_{(i,j) \in \Omega_0} \left( \widehat{D}_{i,j} - D_{\mathrm{ori}, i,j} \right)^2}.
\end{equation}

\textbf{2. Secondary spectrum normalized error ($E_{\mathrm{sec}}$):} 
We compute the RMSE between the linear powers of the inpainted and original secondary spectra over all Fourier pixels, normalized by the original DC power $S_{\mathrm{ori}, (0,0)}$, and report its base 10 logarithm:
\begin{equation} \label{eq:metric_sec}
E_{\mathrm{sec}} = \log_{10} \left( \frac{1}{S_{\mathrm{ori}, (0,0)}} \sqrt{\frac{1}{N_\nu N_t} \sum_{\omega} \left( \widehat{S}_{\omega} - S_{\mathrm{ori}, \omega} \right)^2} \right).
\end{equation}

\textbf{3. Secondary spectrum noise ($P_{\mathrm{noise}}$):} 
To measure the background power, we estimate the mean linear power in the four corner windows of the centered secondary spectrum. Let $\Omega_c$ denote the set of Fourier pixels in these four corners (each spanning one quarter of the full side length along each axis). This background power is expressed in decibels (dB) relative to the DC power of the original secondary spectrum ${S_{\mathrm{ori}, (0,0)}}$:
\begin{equation} \label{eq:metric_noise}
P_{\mathrm{noise}} = 10 \log_{10} \left( \frac{1}{{S_{\mathrm{ori}, (0,0)}}} \frac{1}{|\Omega_c|} \sum_{\omega \in \Omega_c} \widehat{S}_{\omega} \right).
\end{equation}

For $\text{nRMSE}_{\mathrm{dyn}}$ and $E_{\mathrm{sec}}$, lower values indicate closer agreement with the original data. For $P_{\mathrm{noise}}$, lower values represent a cleaner background with less mask-induced leakage.

DT-FISTA gives the lowest inpainting errors in the dynamic and secondary spectra for both tested patches. Figure~\ref{fig:patch_comparison} and Table~\ref{tab:patch_metrics} show that $\mathrm{nRMSE}_{\mathrm{dyn}}$ decreases from 0.297 with Wiener inpainting to 0.188 with DT-FISTA for PSR~B1237+25 (37\%), and from 0.363 to 0.169 for PSR~B2021+51 (53\%). The $E_{\mathrm{sec}}$ values also decrease. Mean-value inpainting has the largest errors and prominent spectral leakage. The regions inpainted by the Wiener method in Figure~\ref{fig:patch_comparison} (c, k) show visible boundary discontinuities. A likely contributor is distortion of the secondary spectrum by the mask, which affects the estimated signal and noise covariances. DT-FISTA's schedule of decreasing thresholds is designed to reduce the influence of weak leakage during the early iterations.

The measured accuracy gain is accompanied by a reduction in runtime by a factor of about 8,000-10,000 and a reduction in memory footprint by a factor of about 4,000 compared to the dense Wiener implementation. DT-FISTA takes 1.7 s and 0.015 GiB per patch, versus 14,263.9-16,568.4~s and 60.4 GiB for Wiener reconstruction (Table~\ref{tab:patch_metrics}).

\begin{figure*}[tp]
\centering
\includegraphics[width=0.95\textwidth]{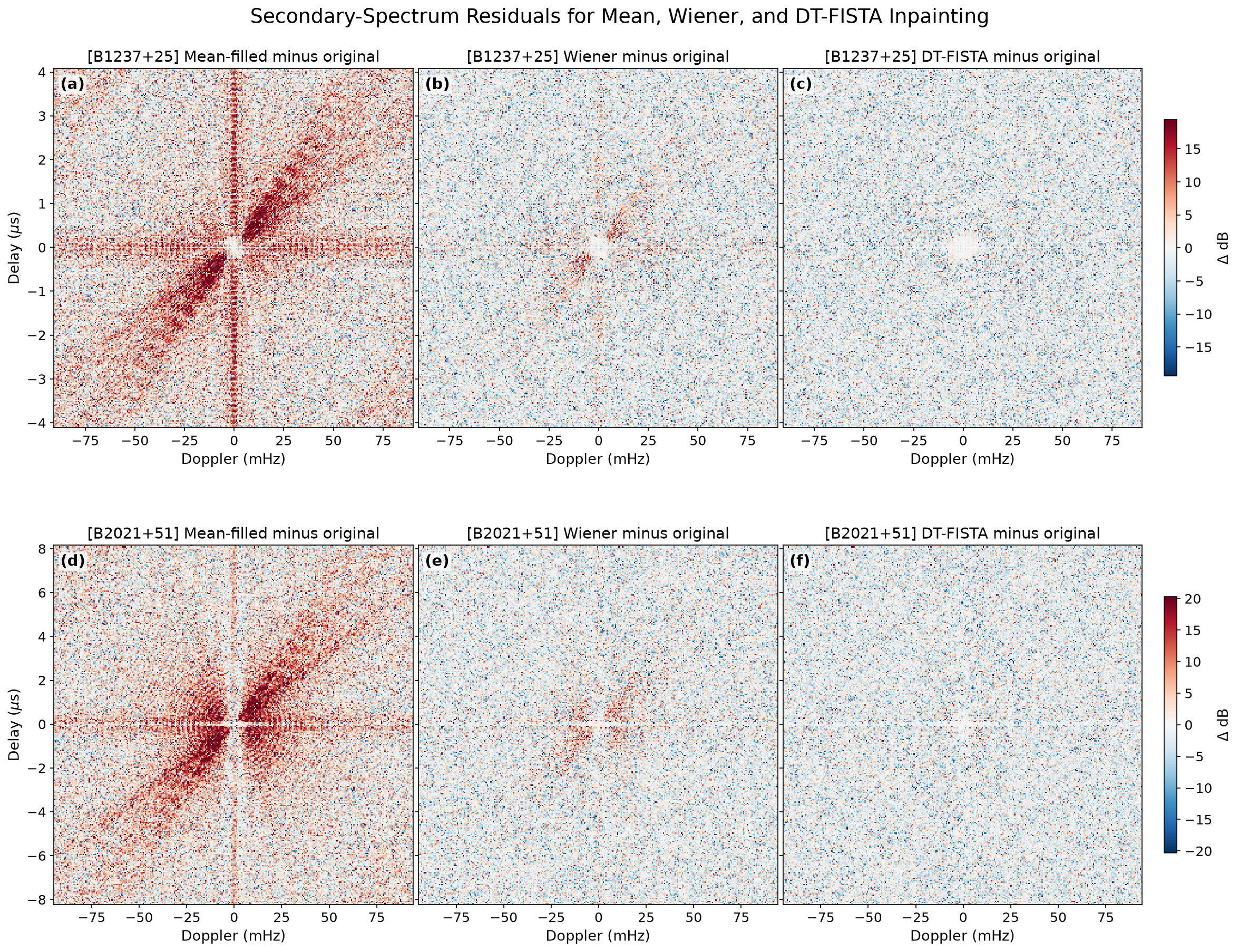}
\caption{Residual maps show where the Fourier power after inpainting differs from the reference. Secondary spectrum residuals are computed by subtracting the original dB-domain spectrum from each post-inpainting dB-domain spectrum. The panels are intended only to visualize the residual structure; the quantitative error is defined in Section~\ref{subsec:comparison}.}
\label{fig:secondary_residuals}
\end{figure*}

\renewcommand{\arraystretch}{1.1}
\begin{table*}[tp]
\centering
\caption{Inpainting errors and resource use: DT-FISTA versus dense Wiener--CG.}
\label{tab:patch_metrics}
\smallskip
\footnotesize
\setlength{\tabcolsep}{7pt}
\renewcommand{\arraystretch}{1.2}

\begin{tabular}{@{}llccccc@{}}
\toprule

Pulsar & Method & \makecell{$\mathrm{nRMSE}_{\mathrm{dyn}}$} & \makecell{$E_{\mathrm{sec}}$} & \makecell{$P_{\mathrm{noise}}$\\(dB)} & \makecell{Time\\(s)} & \makecell{Memory\\(GiB)} \\

\midrule

B1237+25 & Original & \nodata & \nodata & -84.77 & \nodata & \nodata \\
& Mean & 1.020 & -4.10 & -82.40 & \nodata & \nodata \\
& Wiener (dense CG) & 0.297 & -4.62 & -85.49 & 16,568.4& 60.4 \\
& DT-FISTA & 0.188 & -4.83 & -85.65 & 1.72& 0.015 \\

\addlinespace

B2021+51 & Original & \nodata & \nodata & -77.98 & \nodata & \nodata \\
& Mean & 1.249 & -3.64 & -75.94 & \nodata & \nodata \\
& Wiener (dense CG) & 0.363 & -4.04 & -78.66 & 14,148.0& 60.4 \\
& DT-FISTA & 0.169 & -4.73 & -78.86 & 1.72& 0.015 \\ 

\bottomrule
\end{tabular}

\smallskip
\par\raggedright\footnotesize
\textit{Note.} $\mathrm{nRMSE}_{\mathrm{dyn}}$, $E_{\mathrm{sec}}$, and $P_{\mathrm{noise}}$ are defined in Section~\ref{subsec:comparison}. Time is the elapsed time required for an inpainting run (s). Memory is the peak additional resident memory used during the run relative to its pre-run baseline (GiB). The benchmarks were run on Intel Xeon Gold 6230 processors at 2.10 GHz, with 80 logical processors and one numerical thread per inpainting run.
\end{table*}

The secondary spectra obtained after Wiener and DT-FISTA inpainting have $P_{\mathrm{noise}}$ values below those of the original patches. This small decrease occurs because the missing regions are populated by smoother values. Figure \ref{fig:secondary_residuals} shows how the secondary spectra differ. The residuals are computed by subtracting the original dB-domain power from the post-inpainting dB-domain power. Quantitative $E_{\mathrm{sec}}$ is described in Section~\ref{subsec:comparison}.

\begin{figure*}[tp]
\centering
\includegraphics[width=0.95\textwidth]{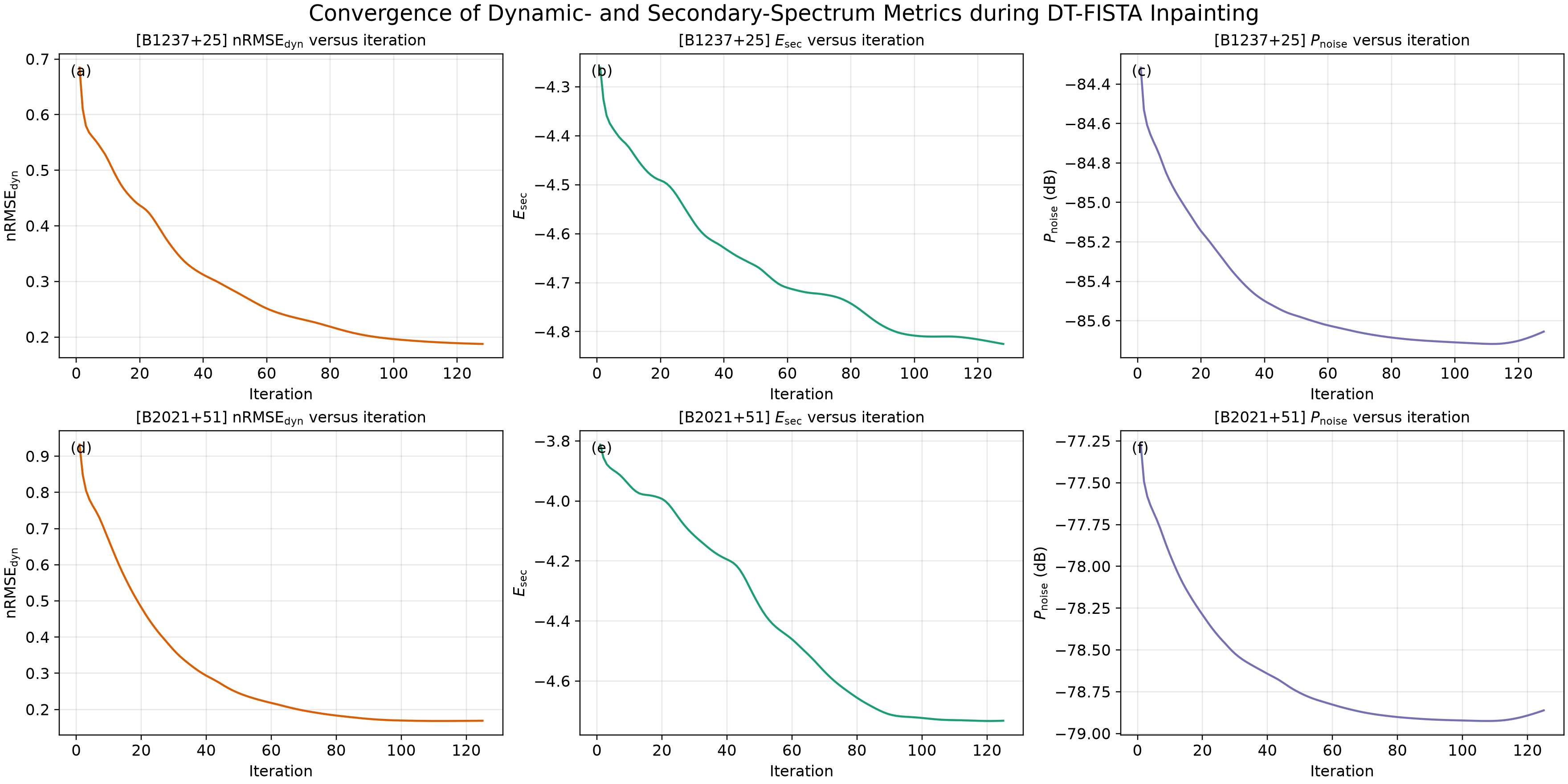}
\caption{Inpainting errors flatten near the empirical stopping scale. Evolution of $\mathrm{nRMSE}_{\mathrm{dyn}}$ (left), $E_{\mathrm{sec}}$ (middle), and $P_{\mathrm{noise}}$ (right) during DT-FISTA inpainting. The top and bottom rows correspond to PSR~B1237+25 and PSR~B2021+51, respectively. The metrics are defined in Section~\ref{subsec:comparison}.}
\label{fig:confista_convergence}
\end{figure*}

Most of the measured error reduction is achieved before the threshold reaches the empirical background scale. Figure~\ref{fig:confista_convergence} follows $\mathrm{nRMSE}_{\mathrm{dyn}}$, $E_{\mathrm{sec}}$, and $P_{\mathrm{noise}}$ during the DT-FISTA iteration sequence. The stopping rule is reached after 128 iterations for PSR~B1237+25 and 125 for PSR~B2021+51. Both error curves decrease and flatten near the median Fourier magnitude. $P_{\mathrm{noise}}$ reaches a minimum shortly before termination and then rises slightly, consistent with admitting coefficients of lower amplitude containing more noise. These two tests support stopping near this scale.

\subsection{Scope and Limitations} \label{subsec:limitations}

The method is most useful when the signal is approximately sparse or compressible in the Fourier domain, and the retained data constrain the missing structure. Inpainting is not guaranteed for diffuse power distributions, large masks, or retained data contaminated by unflagged RFI. Since retained values are enforced exactly, their noise is also retained and can propagate into masked regions.

\begin{figure*}[tp]
\centering
\includegraphics[width=0.95\textwidth]{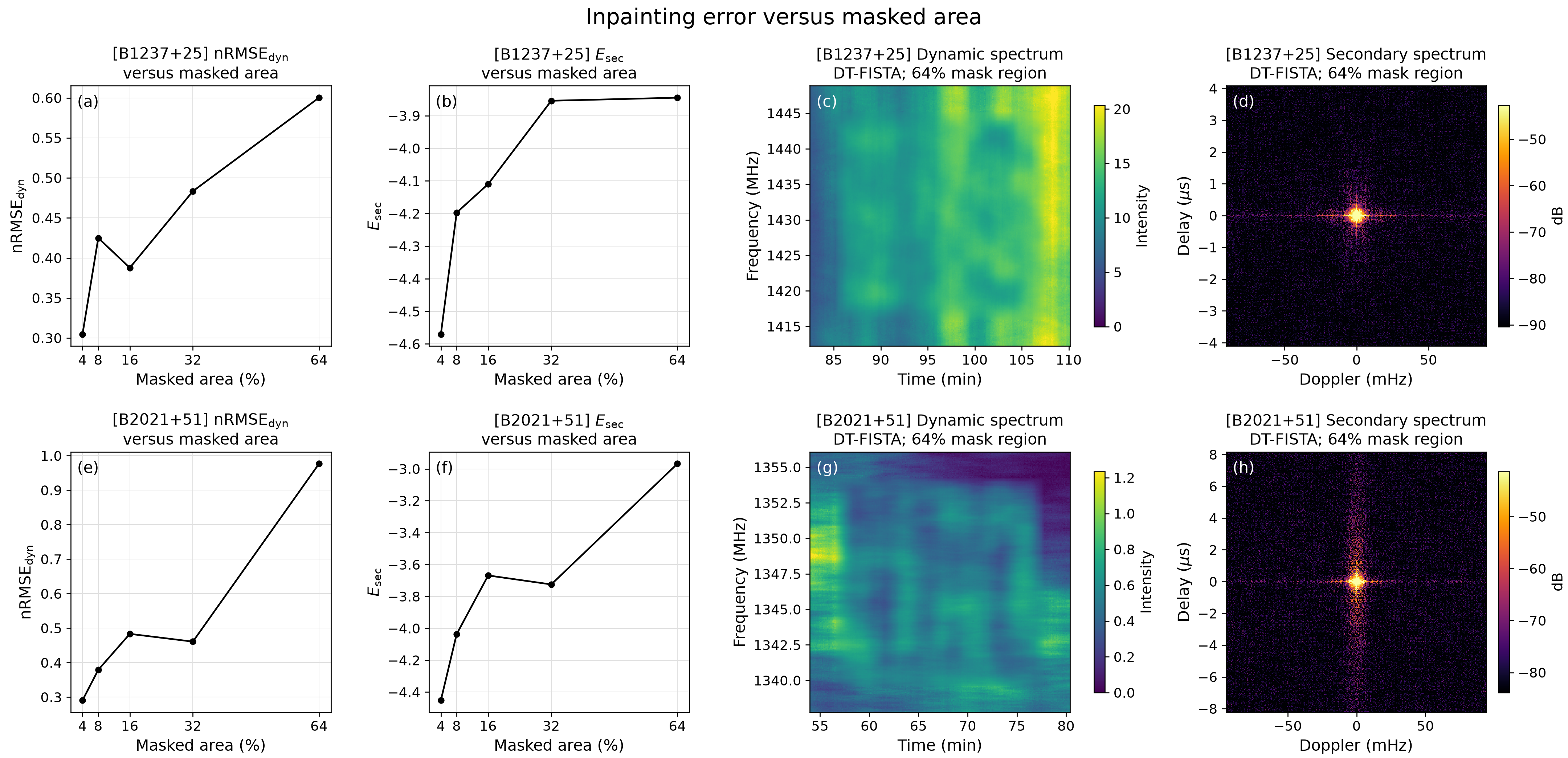}
\caption{Inpainting error and remaining structure as square masks grow. The rows show B1237+25 and B2021+51. The columns show $\mathrm{nRMSE}_{\mathrm{dyn}}$, $E_{\mathrm{sec}}$, and the dynamic spectra inpainted by DT-FISTA and their corresponding secondary spectra for a 64\% mask. Curves use mask areas of 4\%, 8\%, 16\%, 32\%, and 64\%. Secondary spectrum power is in dB relative to DC. Metrics and reference spectra are given in Section~\ref{subsec:comparison} and Figure~\ref{fig:patch_comparison}.}
\label{fig:hole_size}
\end{figure*}

Large contiguous masks leave less information with which to constrain the inpainted estimates. We tested central square masks covering 4\%, 8\%, 16\%, 32\%, and 64\% of each of the two reference patches used in Section~\ref{subsec:patch_results}. Figure~\ref{fig:hole_size} shows the inpainting errors in the dynamic and secondary spectra defined in Section~\ref{subsec:comparison}. Between the smallest and largest masks, $\mathrm{nRMSE}_{\mathrm{dyn}}$ rises from 0.305 to 0.600 for B1237+25 and from 0.291 to 0.977 for B2021+51. The trend is not strictly monotonic for either pulsar. At 64\% mask area, mean-value filling gives $\mathrm{nRMSE}_{\mathrm{dyn}}=0.899$ for B1237+25 and 1.172 for B2021+51, whereas DT-FISTA gives 0.600 and 0.977, respectively, reducing the dynamic-spectrum error by 33\% and 17\%. Thus, DT-FISTA remains more accurate than mean-value filling even when more than half of the pixels are masked.

\begin{figure}[tp]
\centering
\includegraphics[width=\columnwidth]{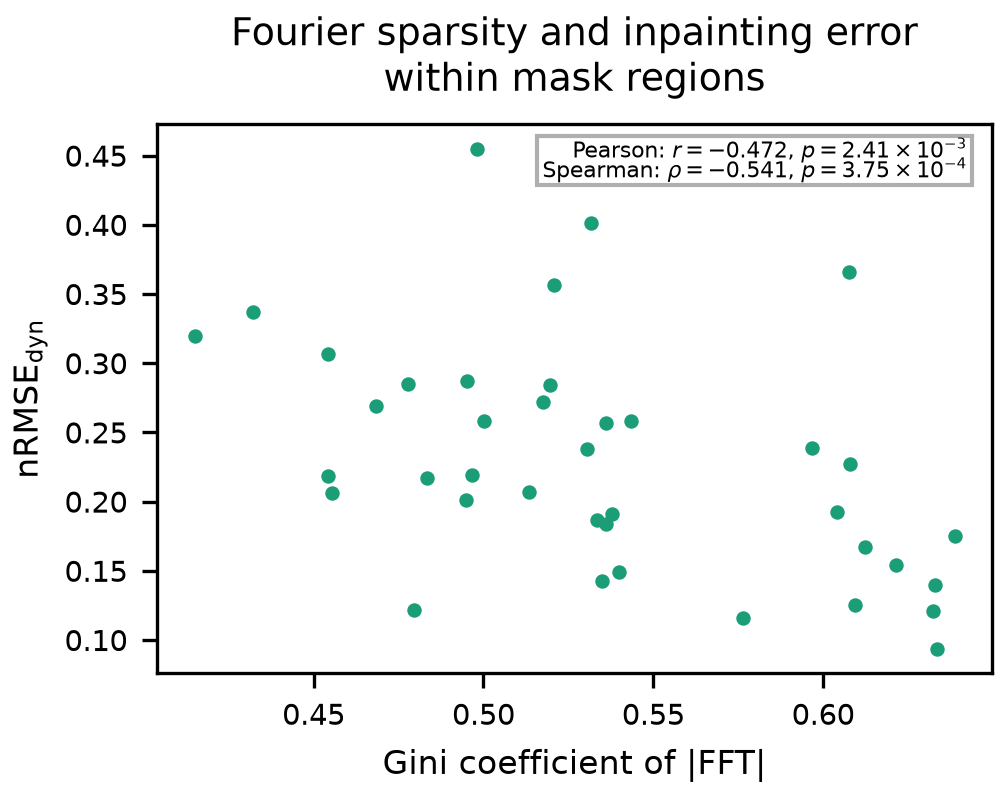}
\caption{Fourier sparsity or compressibility and inpainting error within artificial masks across 39 non-overlapping $300\times300$ patches of B2021+51.}
\label{fig:gini_rmse}
\end{figure}

Sparsity or compressibility in the Fourier domain is also associated with inpainting quality. We selected $300\times300$ patches from B2021+51 whose originally flagged fraction is strictly below 0.5\%, linearly interpolated those flagged samples, and then imposed the same square, circular, and irregular artificial masks as in Section \ref{subsec:patch_results}, covering 15\% in total. We obtained 39 non-overlapping patches and inpainted all of them with the same DT-FISTA parameters. We excluded B1237+25 because no $300\times300$ patch satisfies this stricter mask criterion. We characterize this sparsity or compressibility using the Gini coefficient of the Fourier magnitudes of each unmasked reference patch (before the artificial masks are imposed). The Gini coefficient measures how unevenly values are distributed. A value approaching one indicates that most amplitude is concentrated in a small fraction of coefficients, corresponding to greater sparsity or compressibility. A value approaching zero indicates more uniform amplitudes and thus less sparsity. Figure~\ref{fig:gini_rmse} plots this coefficient against $\mathrm{nRMSE}_{\mathrm{dyn}}$ within the masks.

For these 39 B2021+51 patches, larger Gini coefficients are associated with lower errors. The Pearson product--moment correlation coefficient is $r=-0.472$ with a two sided $p$ value of $2.41\times10^{-3}$, while the Spearman rank correlation coefficient is $\rho=-0.541$ with a two sided $p$ value of $3.75\times10^{-4}$. Therefore, DT-FISTA is most effective for data with concentrated Fourier power, while highly diffuse distributions yield lower inpainting accuracy.

\section{Conclusions} \label{sec:conclusion}

In this work, we introduced DT-FISTA, an inpainting method for recovering missing data by exploiting sparse Fourier power in the observed data. By combining Fourier domain continuation with a decreasing threshold schedule and Nesterov momentum, DT-FISTA progressively recovers masked scintillation structures from coarse to fine scales while suppressing mask-induced spectral leakage.

In controlled tests on $300\times300$ dynamic-spectrum patches, DT-FISTA achieved lower errors in both the dynamic and secondary spectra than the dense Wiener implementation used for comparison, while reducing runtime and memory use by factors of approximately $10^4$ and $4\times10^3$, respectively. Its per-iteration cost of $\mathcal{O}(N\log N)$ and memory requirement of $\mathcal{O}(N)$ make full-resolution processing of large observations feasible.

For scintillation studies, DT-FISTA provides a preprocessing step that mitigates the artifacts introduced by RFI excision in secondary spectra. By suppressing mask-induced spectral leakage and improving the detectability and characterization of faint scintillation arcs, it can enable more reliable constraints on the physical properties of interstellar scattering screens from full-resolution observations.

\section{Acknowledgments}

This work is supported by the National Natural Science Foundation of China under grant No. 12373025. Yuanshang Huang is supported by Scientific Research and Innovation Project of Postgraduate Students in the Academic Degree of Yunnan University (Grant No. KC-252513932) and the Project of Yunnan Provincial Department of Education Science Research Fund (Grant No. 2026Y0175). 

Yuanshang Huang is grateful to Tianyue Li (China Academy of Engineering Physics) for a careful reading of the manuscript and valuable suggestions that improved the logic and clarity of the background and methodology sections. He also thanks Guang-Xing Li (SWIFAR, Yunnan University) for helpful discussions of the mathematical foundations of the methods, and Panpan Huang (SWIFAR, Yunnan University) for suggestions on the paper's structure and constructive critiques of the methodology.

This work made use of the data from FAST (Five-hundred-meter Aperture Spherical radio Telescope). FAST is a Chinese national mega-science facility, operated by National Astronomical Observatories, Chinese Academy of Sciences.

\facility{FAST:500m}.

\software{SciPy \citep{2020SciPy-NMeth}, Matplotlib \citep{Hunter:2007} and NumPy \citep{harris2020array}.}

\section{Data Availability} \label{sec:dataavailability}

The code and datasets used in this work are publicly available at \url{https://github.com/HuangYuanshang/DT-FISTA}.

\appendix
\section{Convexity of the Sparse Inpainting Reference Problem}\label{app:convexity}

For any two arrays $D_1,D_2$, write $A_\omega=\mathcal{F}(D_1)_\omega$ and $B_\omega=\mathcal{F}(D_2)_\omega$. For $0\le a\le1$,
\begin{equation} \label{eq:convexity}
\begin{aligned}
&\mathcal{R}(aD_1+(1-a)D_2)\\
&\quad=\sum_{\omega\ne(0,0)}|aA_\omega+(1-a)B_\omega|\\
&\quad\le\sum_{\omega\ne(0,0)}\bigl(a|A_\omega|+(1-a)|B_\omega|\bigr)\\
&\quad=a\mathcal{R}(D_1)+(1-a)\mathcal{R}(D_2).
\end{aligned}
\end{equation}
The first equality uses Fourier linearity. The inequality applies $|u+v|\le|u|+|v|$ to each coefficient, with $a,1-a\ge0$ allowing their magnitudes to factor out. The final bound is the defining inequality for a convex function.

The data consistency set is also convex. For any $D_1,D_2\in\mathcal{C}_{\mathrm{data}}$ and any real $a$,
\begin{equation}
\label{eq:cdata_convexity}
\begin{aligned}
&M\odot\bigl(aD_1+(1-a)D_2\bigr)\\
&\quad=a(M\odot D_1)+(1-a)(M\odot D_2)\\
&\quad=a(M\odot D_{\mathrm{con}}^{(0)})+(1-a)(M\odot D_{\mathrm{con}}^{(0)})\\
&\quad=M\odot D_{\mathrm{con}}^{(0)}.
\end{aligned}
\end{equation}
Equation~(\ref{eq:cdata_convexity}) shows that $\mathcal{C}_{\mathrm{data}}$ is affine: every affine combination of feasible arrays remains feasible. Restricting to $0\le a\le1$ proves convexity, since every line segment between feasible arrays lies in the set. Together with Equation~(\ref{eq:convexity}), this makes Equation~(\ref{eq:bp_reference}) a convex reference model \citep{chen98}.

\section{Edge Repair} \label{app:edge_repair}

After the first pass, we divide the inpainted estimate by the Hanning window and restore the retained mean, as in Equation~(\ref{eq:output}). Small window values can amplify errors in masked regions near the boundaries. We therefore run a second DT-FISTA pass without a Hanning window to inpaint the originally masked pixels near the edges, treating all other pixels, including interior regions inpainted in the first pass, as retained data. This pass subtracts the mean of its retained samples, initializes the target masked pixels to zero, applies the same threshold and stopping rules, and restores the mean afterward. Only the target edge pixels receive the new estimates. Figure~\ref{fig:full_reconstruction} uses this repair for the full observations.

If only a secondary spectrum with low background noise is needed, compute $S(D_{\mathrm{con}}^{(K)})$ directly from the final windowed estimate in Figure~\ref{fig:confista_flowchart} (o). Division by the window, restoration of the mean, and edge repair are then unnecessary. Retaining the window avoids amplification of boundary errors.

\bibliography{DT-FISTA}{}
\bibliographystyle{aasjournalv7}

\end{document}